\documentclass[]{iopart}
\usepackage{graphicx}

\DeclareGraphicsExtensions{.eps,.ps,.jpg,.gif,.tif,.png}

\usepackage{array}

\begin{document}
\title[v3.3; 11Aug2010 ; SST]{A scalable readout system for a superconducting adiabatic quantum optimization system}

\author{
  A.~J.~Berkley, M.~W.~Johnson, P.~Bunyk, R.~Harris,
  J.~Johansson, T.~Lanting, E.~Ladizinsky, E.~Tolkacheva,
  M.~H.~S.~Amin and G.~Rose}
  \address{All authors are with D-Wave
    Systems Inc., 100-4401 Still Creek Drive, Burnaby, BC V5C 6G9
    Canada}
  \ead{ajb@dwavesys.com} 

\begin{abstract}
  We have designed, fabricated and tested an XY-addressable readout
  system that is specifically tailored for the reading of
  superconducting flux qubits in an integrated circuit that could
  enable adiabatic quantum optimization.  In such a system, the flux
  qubits only need to be read at the end of an adiabatic evolution
  when quantum mechanical tunneling has been suppressed, thus
  simplifying many aspects of the readout process.  The readout
  architecture for an $N$-qubit adiabatic quantum optimization system
  comprises $N$ hysteretic dc SQUIDs and $N$ rf SQUID latches
  controlled by $2\sqrt{N} + 2$ bias lines.  The latching elements are
  coupled to the qubits and the dc SQUIDs are then coupled to the
  latching elements.  This readout scheme provides two key advantages:
  First, the latching elements provide exceptional flux sensitivity
  that significantly exceeds what may be achieved by directly coupling
  the flux qubits to the dc SQUIDs using a practical mutual
  inductance.  Second, the states of the latching elements are robust
  against the influence of ac currents generated by the switching of
  the hysteretic dc SQUIDs, thus allowing one to interrogate the
  latching elements repeatedly so as to mitigate the effects of
  stochastic switching of the dc SQUIDs.  We demonstrate that it is
  possible to achieve single qubit read error rates of $<10^{-6}$ with this
  readout scheme.  We have characterized the system-level performance
  of a 128-qubit readout system and have measured a readout error probability of
  $8\times10^{-5}$ in the presence of optimal latching element bias conditions.
\end{abstract}

\maketitle

\section{Introduction}\label{intro}

Many proposals exist for how to build superconducting quantum
computing systems \cite{bockoherrfeldman,sqcreview1}, each with their
own unique challenges.  One such proposal is to enable adiabatic
quantum optimization (AQO) using networks of inductively coupled rf
SQUID flux qubits \cite{kaminskylloyd,farhiaqc,grajcaraqc}.  In this
version of AQO, one first maps an optimization problem of interest
onto that of minimizing a classical energy function $V_{IS}$, describing a set of binary
variables $i=0...N$, each having a possible state $s_{z,i}
= \pm 1$, with applied local biases $\epsilon_i$ and pairwise couplings $J_{ij}$:
\begin{equation}
V_{IS} = \sum_{i=0}^{N} \epsilon_i {\mathbf s_{z,i}} + \sum_{i \neq j} J_{ij} {\mathbf s_{z,i} s_{z,j}}
\end{equation}
One then embeds this potential into a physical
quantum Ising spin system with a controllable transverse
magnetic field term $T = \sum_{i=0}^{N} \Delta_q {\mathbf \sigma_{x,i}}$.  This physical system is described by the quantum Ising spin Hamiltonian $H_{IS}$:
\begin{equation}
H_{IS} = T + V_{IS} = \sum_{i=0}^{N} \Delta_q \sigma_{x,i} + \sum_{i=0}^{N} \epsilon_i \sigma_{z,i} + \sum_{i \neq j} J_{ij} \sigma_{z,i} \sigma_{z,j}
\end{equation}
where $\sigma_{x,i}$ and $\sigma_{z,i}$ are Pauli matrices acting on the $i$th spin.

The AQO algorithm proceeds by first initializing the processor in a
state wherein all single spin tunneling energies $\Delta_q$
(proportional to transverse magnetic field) are much larger than the
energy scales involved in the problem of interest.  In this case, all
spins will readily relax into their ground states.  Thereafter, the
AQO algorithm for finding an optimal solution proceeds by smoothly
decreasing $\Delta_q$ in time from a large value until it is much less
than the energy scales involved in the problem of interest.  If the
resultant evolution of the state of the processor is adiabatic, then
its final state will encode an optimal solution to the problem of
interest.

Details of how an Ising spin system with effective transverse magnetic
field is implemented using superconducting flux qubits can be found in
\cite{PhysRevB.82.024511}.  Briefly, rf SQUID flux qubits are used to
represent localized spins, with the spin state being encoded directly
in the circulating current of the rf SQUID\cite{rhccjj}.  Each rf
SQUID has a local flux bias providing the effective applied
longitudinal field term $\epsilon_i$.  The effective transverse
magnetic field at each spin is realized via the tunneling between the
two circulating current states as provided by the presence of at least
one Josephson junction in the rf SQUID.  Tunable inductive coupling
between the rf SQUID flux qubits\cite{PhysRevB.80.052506} provide the
effective spin-spin coupling terms in the Ising Spin Hamiltonian.

The challenge that is addressed in this article is that of reading the
final state of an AQO processor.  Note that at the end of the AQO
algorithm $\Delta_q$ is negligible, thus naturally terminating the
evolution in a state that is diagonal in the qubits' flux bases.  This
is in contrast to a general purpose gate model (GM) quantum
information processor in which the final state of the processor could
be a superposition state due to appreciable $\Delta_q$.  The method
presented in this paper for distinguishing the flux in a large number
of devices should also find use in classical superconducting logic
circuits.

Several methods for reading the state of a flux qubit in its flux
basis have been reported upon in the literature to date.  One approach
is to inductively couple a qubit to a hysteretic dc SQUID and then
interrogate the latter device using either a current bias ramp
\cite{dcsquidreadout0,dcsquidreadout1} or carefully crafted current
pulse \cite{dcSQUIDpulse}.  Either way, the objective is to discern
the current bias at which the dc SQUID switches into the voltage
state, which depends upon the flux imparted to the dc SQUID by the
qubit.  One key advantage to this readout mechanism is its physical
size: dc SQUIDs can be made relatively small, thus providing
favourable prospects for utilizing such a readout mechanism in
many-qubit processors.
Furthermore, as shown in this article, one can design scalable readout
biasing architectures based upon dc SQUIDs that make economical use of
a limited number of external bias controls.  On the other hand, while
hysteretic dc SQUIDs were used in early GM quantum computation
experiments, they have fallen out of favour due to on-chip heating and
back-action on the qubits.  The first of these drawbacks could be
remedied, in principle, by better thermal design.  The second drawback
is less of an issue for AQO as the quantum computation itself
naturally localizes the state of each qubit by making $\Delta_q$
negligible by the end of the computation, thus inhibiting transitions
within the qubit.  As such, hysteretic dc SQUIDs could still be of use
in designing scalable readout architectures for AQO processors.  Note
that readout architectures based upon scalable dispersive techniques
\cite{dispersivereadout0,dispersivereadout1}, which may
yet prove to be essential in the development of GM processors, take up
significantly more physical space than those based upon solely upon hysteretic dc
SQUIDs.  Nonetheless, future developments may eventually make
dispersive readout methods an attractive option for an AQO.

While the arguments presented above suggest that the relatively simple
hysteretic dc SQUID readout may be useful in the
development of large scalable AQO processors, our research has
revealed one more drawback of such an approach.  It is well known that
unshunted dc SQUIDs can resonate when switched into the voltage state,
thus acting as an on-chip source of microwave radiation.  While this
feature has been exploited to resonantly excite flux qubits in novel
quantum mechanics experiments \cite{LukensSuperposition}, this
behavior would be highly undesirable in a future functional quantum
information processor.  We have observed that this behavior is
particularly problematic in dense superconducting circuits as the
radiation generated by a dc SQUID can resonantly drive multiple flux
qubits within its vicinity, thus altering their final states.  Part of
this issue may simply be an unfortunate choice of rf SQUID parameters:  The
design pressures in an AQO processor tend to favour flux qubits with
large geometric inductances so as to facilitate multiple inter-qubit
inductive couplings.  Consequently, in order to achieve appreciable
$\Delta_q$ during the AQO algorithm, one must lower the designed critical current of the rf
SQUID, thus yielding a flux qubit with a relatively low tunnel
barrier\footnote{For comparison, typical barrier heights in rf SQUID
  phase qubits reported on in the literature are $\sim 1$ THz, as compared to $\sim 500$ GHz for our flux qubits.  When we design our qubits to have tunnel barrier heights that are in excess of $\sim 1$ THz, we do not observe resonant activation.  Unfortunately, such flux qubits are not optimal for use in AQO as problems related to Josephson junction critical current variation are accentuated.}
over which radiation from dc SQUIDs can drive resonant activation.
This is clearly unacceptable, as the solution to the optimization
problem that has been posed to the processor is then corrupted by the
act of reading.  The novel approach that we have implemented to remedy
this problem is to insert a quantum flux parametron (QFP)
\cite{likharevqfp,gotoqfp} between each qubit and its dedicated dc
SQUID.  The QFP is an rf SQUID that possesses a small inductance, a
large capacitance and a very large critical current.  Consequently,
the QFP possesses a substantially larger tunnel barrier as compared to
the flux qubit, which makes its state robust against
high frequency radiation emanating from dc SQUIDs in the voltage state.
In our architecture, the QFP is used to quietly latch the final state
of a given flux qubit with minimal back-action.  Thereafter, the final
state of the QFP is read using a hysteretic dc SQUID.  Thus, the QFP
and dc SQUID combination provides a viable route to a scalable readout
architecture that is suitable for an AQO processor.

\section{Readout Circuit Design}\label{design}

\begin{figure}
\centering
\includegraphics[width=3in]{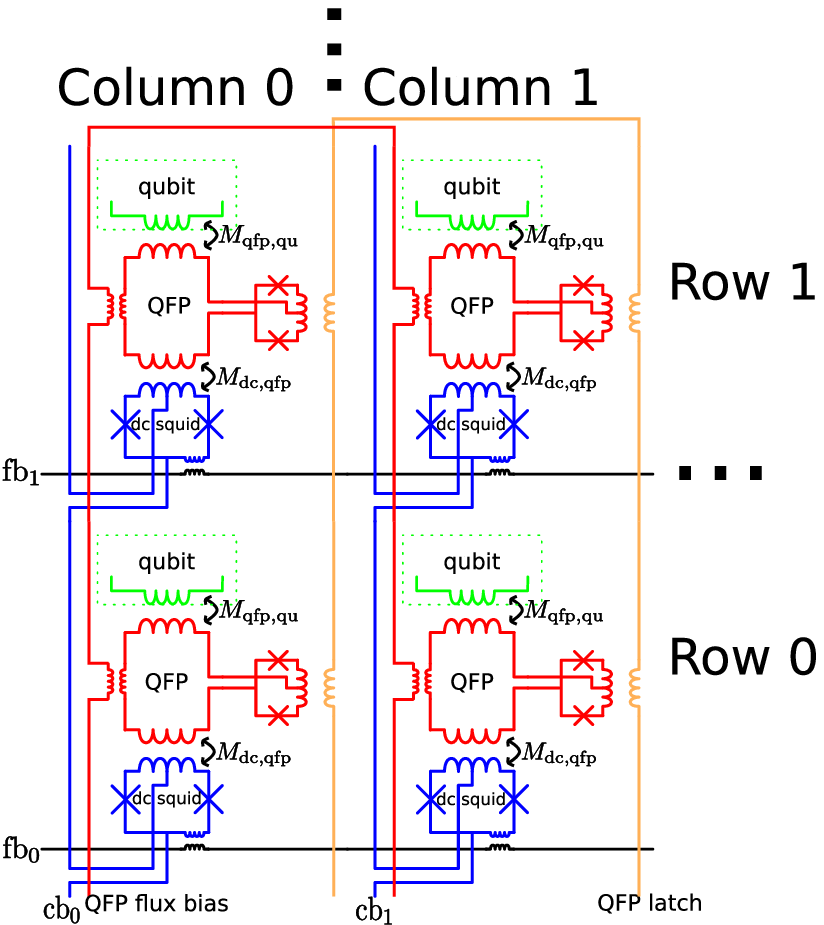}
\caption{XY addressable readout system schematic.  Each dc SQUID has two shared
  control lines: a current bias (cb$_0$, cb$_1$, $...$) and a flux
  bias (fb$_0$, fb$_1$, $...$).  All QFP latches share a single
  activation line (``QFP latch'') and a single flux
  bias line (``QFP flux bias'').  The qubits are discussed in detail
  in Refs.~\cite{rhccjj,mwjpmm}.}
\label{fig:roschematic}
\end{figure}

\begin{figure}
\centering
\includegraphics[width=3in]{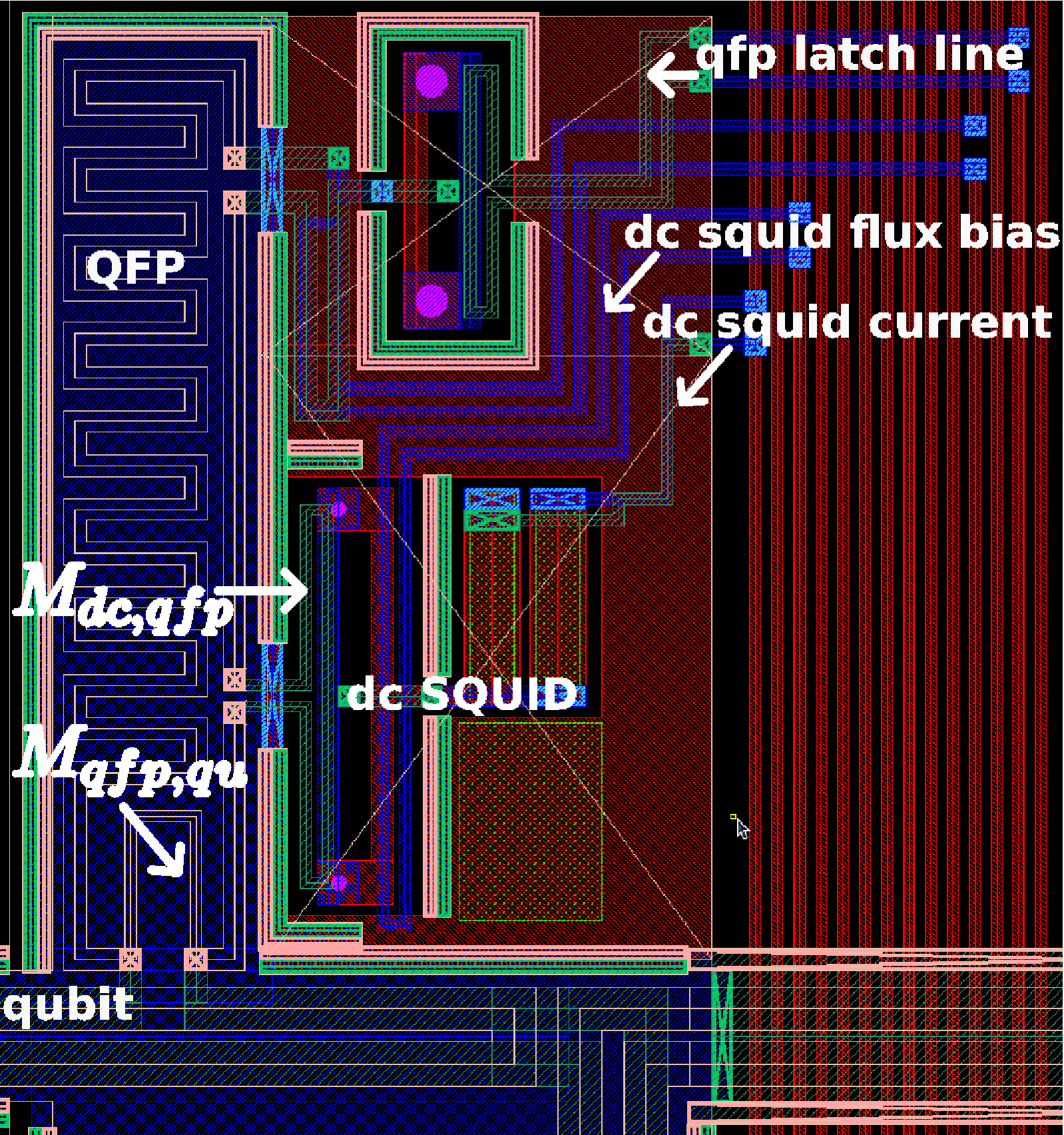}
\vskip0.1in
\includegraphics[width=3in]{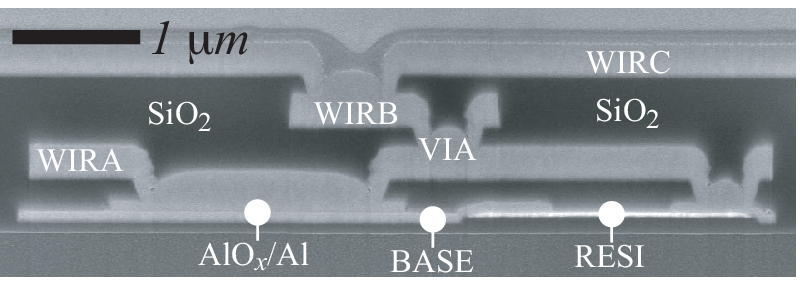}
\caption{(top) Layout of a portion of an XY readout system.  Note that
  all of the elements are shielded with a ground plane.  The region
  shown above is 47x50 $\mu$m$^2$.  There are four Nb metal layers: BASE
  (red), WIRA (blue), WIRB (green diagonal hashed), WIRC (beige), and RESI (green dotted) is a Ti/Pt resistor layer.  Junctions are denoted as AlO$_x$/Al (pink circles) and
  vias are denoted by crossed rectangles.  (bottom) Cross-section of the planarized fabrication process.
  All metal layers have a nominal thickness of 300 nm and the dielectric thickness is 200 nm.}
\label{fig:rolayout}
\end{figure}

Figure~\ref{fig:roschematic} shows a high level schematic of a small portion of an XY-addressable readout
system.  Since this article is focused upon the readout architecture, we have suppressed the details
of the flux qubits and their control circuitry and refer the reader to Refs.~\cite{rhccjj,mwjpmm}
for information concerning those elements.  Each qubit in the circuit
requires two elements for readout: a dc SQUID and a QFP.  Each dc
SQUID has two shared control lines: a current bias (cb$_0$, cb$_1$,
$...$) and a flux bias (fb$_0$, fb$_1$, $...$).  All QFP latches share
a single activation line (``QFP latch'', flux bias referred to as
$\Phi^x_{\mathrm{latch}}$) and a single flux bias line (``QFP flux
bias'', flux bias referred to as $\Phi^x_{\mathrm{qfp}}$).  Let the
inductance and capacitance of the QFP be denoted by $L_{\mathrm{qfp}}$
and $C_{\mathrm{qfp}}$, respectively.  Each QFP is coupled to its
affiliated qubit (dc SQUID) via a mutual inductance
$M_{\mathrm{qfp,qu}}$ ($M_{\mathrm{dc,qfp}}$).  The top panel of
figure~\ref{fig:rolayout} shows the physical layout of a portion of the readout system, as implemented in the chips discussed in this article.   Note that the QFP is galvanically attached to the qubit; this is done purely to reduce the contribution of the readout to the length and inductance of the qubit.  The bottom panel of this figure shows a cross-section of our planarized fabrication process that was used to produce the test circuits reported upon herein.  All results presented in this article were obtained from test circuits containing either 8 or 128 flux qubits, as indicated in the text.

\begin{figure}
\begin{center}
\includegraphics[width=3in]{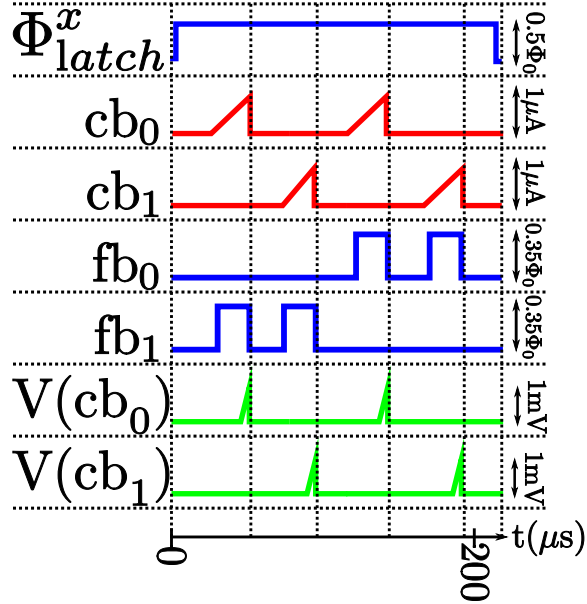}
\end{center}
\caption{Waveforms for operating the 2x2 XY addressable readout depicted in figure~\ref{fig:roschematic}.  cb$_0$ and
  cb$_1$ are current bias lines carrying linear ramps with typical height of 1-2 $\mu$A.  fb$_0$ and
  fb$_1$ are flux bias lines carrying pulse currents of height 1-2 mA corresponding to pulsing from 0.0 to 0.35
  $\Phi_0$ applied to a dc SQUID.  QFP latch line carries roughly 1-2 mA signal varying from 0.5 $\Phi_0$
  (quiescent) to 1 $\Phi_0$ (latched) in the QFP.  V(cb$_x$) schematically depicts the voltage on the current
  bias lines when a single dc SQUID in a column is switched.  The voltage signals shift slightly left and right
  in time depending on the state of the QFP.  The voltage detection is done with a simple threshold
  comparator.}\label{fig:waveforms}
\end{figure}

At a high level, the readout circuit functions as follows: During the
course of an adiabatic quantum computation, all current biases $cb_x$
and flux biases $fb_y$ are set to zero, thus ensuring that the dc
SQUIDs are in the zero voltage state.  In order to decouple the QFPs
from the flux qubits, the QFP latch line is set to provide
$\Phi^x_{\mathrm{latch}}\approx \Phi_0/2$ to all QFPs, thus
suppressing their persistent currents to negligible levels.  Once the
qubits have reached their final states at the end of an adiabatic
quantum computation, the waveform sequence depicted in figure~\ref{fig:waveforms} is applied to the readout circuitry.  First, the QFP latch bias is raised to provide $\Phi^x_{\mathrm{latch}}\approx \Phi_0$ to all QFPs, thus raising the QFP tunnel barriers (see figure~\ref{fig:qfppotentialenergy}), which adiabatically changes the state of each QFP from being the ground state of a monostable potential to being the groundstate of a bistable potential.  The direction of the resultant groundstate persistent current flow about the body of each QFP is determined by the polarity of the flux imparted by the qubit to which each QFP is coupled.  At the end of this linear ramp the height of the QFP tunnel barriers $\Delta U_{\mathrm{qfp}}$ has been raised to a sufficiently high level so as to preclude further dynamics
($\Delta U_{\mathrm{qfp}} \gg \hbar / \sqrt{L_{\mathrm{qfp}} C_{\mathrm{qfp}}} \gg k_b T$, where $T$ is the temperature of the chip), and the final states of the flux qubits have been latched into the QFPs.  Thereafter, a linear current ramp is applied to the bias cb$_x$ whose maximum amplitude is significantly lower than the maximum dc SQUID switching current $I^{\mathrm{sw}}_{\mathrm{max}}$ amongst the population of dc SQUIDs connected in series with cb$_x$, so that only the dc SQUID with a flux bias applied to its fb$_y$ line is triggered (see figure~\ref{fig:dcsquidmodulation}).  The timing of a voltage arising across the dc SQUID is then measured: If the flux arising from the QFP increases (decreases) the total flux applied to the dc SQUID addressed by cb$_x$ and fb$_y$, then switching to the voltage state will happen sooner (later) (see figure 3 and ~\ref{fig:dcsquidmultireads}). This difference in timing reveals the state of the QFP and therefore the state of the qubit.  A more detailed discussion of each of the elements in this readout architecture and their operation is presented below.

\subsection{Qubit signal}
For the purposes of reading the final state of the flux qubits in an AQO processor, the relevant signal is the magnitude of the qubit persistent current when the qubit tunnel barriers have been raised to their maxima.  The flux qubits used in our circuits are of the type described in Ref.~\cite{rhccjj} and had inductances $L_q \sim 300\,$pH, of which $10\,$pH was reserved for readout, and critical currents $I_q^c\sim 2.2\,\mu$A.  During readout, the qubit $\beta_q=2\pi L_qI^c_q/\Phi_0\sim2$, corresponding to a circulating current of roughly $I^q_p \sim 2 \mu A$.  A typical mutual
inductance from the qubit into its corresponding QFP is $M_{\mathrm{qfp},\mathrm{qu}} \sim 5.4$ pH, leading to a qubit signal as seen by the QFP of $2 I^p_q M_{\mathrm{qfp},\mathrm{qu}} \sim 10\,$m$\Phi_0$.

For the qubit parameters cited above, the maximum tunnel barrier height between flux states would be $\Delta U_q/h \sim 460\,$GHz.  The design pressures that led to this choice are discussed in detail in \cite{rhccjj,mwjpmm} and so we only briefly repeat the relevant points herein.  While in theory one could design flux qubits with larger $I_q^c$ (and thus a larger $\Delta U_q$ to provide immunity to any dc SQUID driving signal) both asymmetry in the qubit junctions and finite precision on the qubit analog flux biases would eventually limit circuit performance as an AQO processor.  Furthermore, typical AQO architectures require a substantial number of inter-qubit couplings \cite{kaminskylloyd} which necessitates larger $L_q$ than may be considered appropriate in the study of flux qubits for implementing GM hardware.  As such, the flux qubits in our circuits tend to possess substantially smaller $I_q^c$ in order to realize appreciable single qubit tunneling energies $\Delta_q$ over as broad of a range of annealing parameter as is practical.  Consequently, our flux qubits appear to be more susceptible to resonant activation by dc SQUID switching transients than other flux qubits reported upon in the literature.

\subsection{QFP latch}
\begin{figure}
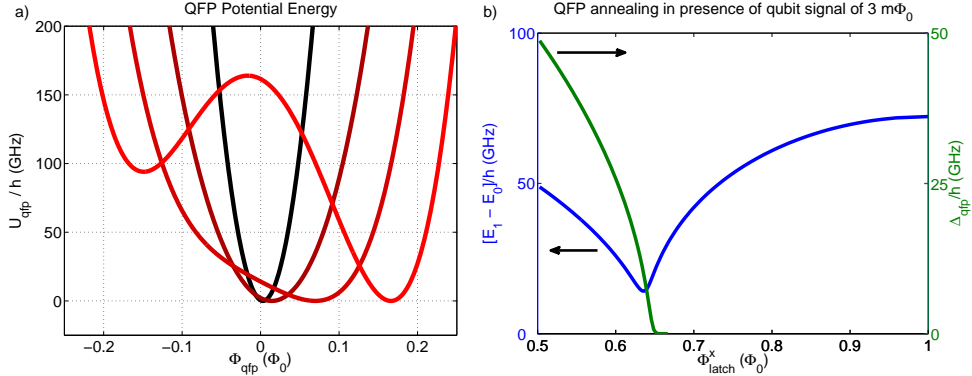

\centering
\includegraphics[height=2.0in]{QFPpotentialenergy.psc}
\includegraphics[height=2.0in]{qfpanneal.psc}
\caption{a) Depiction of the potential energy of the QFP as the CJJ
  bias is raised from $\Phi_0 / 2$ toward $\Phi_0$ during the readout
  process.  Model parameters are $L_{\mathrm{qfp}} = 65$ pH,
  $I^c_{\mathrm{qfp}}=12\,\mu$A, and $C_{\mathrm{qfp}}$ = 160 fF.  The two
  counter-circulating persistent current states correspond to the two
  minima of the bistable (red) potential energy landscape.  The final
  QFP barrier height for $\Phi^x_{\mathrm{latch}}/\Phi_0=1$ is near
  3.5 THz (not shown on this figure).  In the example shown, the qubit
  has been assumed to apply $+3\,$m$\Phi_0$, causing the QFP to latch
  into the right well.  The corresponding $\beta_{\mathrm{qfp}}$
  values of the curves, from monostable (black) to bistable (red) are:
  0, 0.79, 0.99, 1.18.  These are achieved by applying
  $\Phi^x_{\mathrm{latch}}/\Phi_0=0.5$, 0.602, 0.629, and 0.657
  $\Phi_0$, respectively.  b)
  Calculation of the ground to first excited state energy difference, E$_1$-E$_0$, of
  the QFP during annealing and the tunneling energy between them, $\Delta_{\mathrm{qfp}}$.  In this example calculation, the significant applied signal from the
  qubit ensures that  E$_1$-E$_0$ is large.  Transitions between the two flux
  states of the QFP stop at approximately when $\Phi^x_{latch} >
  0.65 \Phi_0$.}
\label{fig:qfppotentialenergy}
\end{figure}

The QFP is a 2-junction rf SQUID, which, in comparison to our flux qubits \cite{rhccjj}, possesses a lower inductance $L_{\mathrm{qfp}} \approx 65\,$pH and a substantially higher critical current $I^c_{\mathrm{qfp}} \approx 12\,\mu$A.  Consequently, the former device has a maximum $\beta_{\mathrm{qfp}} = 2\pi L_{qfp}I^c_{\mathrm{qfp}}/\Phi_0 = 2.37$, a maximum persistent current $I^p_{\mathrm{qfp}}\approx 10.5\,\mu$A, and a maximum tunnel barrier between its two persistent current states on the order of $\Delta U_{\mathrm{qfp}}\sim 3.4\,$THz (compare with 0.46 THz for the qubit).  The effective 1-dimensional potential energy of a QFP with identical Josephson junctions can be written as
\begin{eqnarray}
U_{\mathrm{qfp}} & = & -\frac{\Phi_0 I^c_{\mathrm{qfp}}}{2\pi}\cos\left(\frac{\pi\Phi^x_{\mathrm{latch}}}{\Phi_0}\right)\cos(\frac{2\pi\Phi_{\mathrm{qfp}}}{\Phi_0}) \nonumber\\
  & & + \frac{1}{2L_{\mathrm{qfp}}}\left(\Phi_{\mathrm{qfp}}-\Phi^x_{\mathrm{qfp}}\right)^2
\end{eqnarray}
where $\phi\equiv2\pi\Phi_{\mathrm{qfp}}/\Phi_0$ is the mean phase
across the two Josephson junctions.  Note that in tuning
$\Phi^x_{\mathrm{latch}}$ from $\Phi_0/2$ to $\Phi_0$ that
$U_{\mathrm{qfp}}(\phi)$ morphs from being monostable to being
bistable (see figure~\ref{fig:qfppotentialenergy}).  Raising $\Phi^x_{\mathrm{latch}}$ slowly adiabatically latches the state of each flux qubit into the QFP to which it is coupled.  Adiabaticity is guaranteed here due to the large signal from the qubit and the relatively low bandwidth of the QFP latch line which has a rise time of approximately 1 $\mu s$.  Adiabaticity is violated during the latching phase only if there is no signal applied to the QFP, as the two QFP circulating current states are then degenerate in energy in the large $\beta_{\mathrm{qfp}}$ limit.  Note that if the qubit state were to be unchanged after dc SQUID switching, then the QFP could be operated fully reversibly: the QFP could then be reset in the presence of the signal in which it was initially latched, so the whole cycle of barrier raising and lowering would then be adiabatic.

One challenge in designing QFPs (or any 2-junction rf SQUID) is that junction asymmetry leads to
a $\Phi^x_{\mathrm{latch}}$-dependent effective flux offset \cite{rhccjj}.  This offset limits how small one can make $M_{\mathrm{qfp,qu}}$, as one would like the bimodal qubit signal to straddle any such flux offset for values of $\Phi^x_{\mathrm{latch}}$ about which the QFP becomes bistable.  Assuming a typical fabrication spread of 1\% in junction critical currents and a maximum $\beta_{\mathrm{qfp}}\sim 2.5$, this offset could be up to $4\,$m$\Phi_0$.  For this reason, we had chosen $M_{\mathrm{qfp,qu}}$ such that $2I_q^pM_{\mathrm{qfp,qu}}\sim 10\,$m$\Phi_0$. 

There are two other features of the QFP latch that make it useful beyond the qubit state readout method described thusfar. {\it Linear detection:}
By biasing the QFP such that $\beta_{\mathrm qfp} \sim 1$, it can be used as a preamplifier acting on the
signal from the qubit.  Realistically, achieving appreciable net flux gain into the dc SQUID
requires careful tuning of the QFP and dc SQUID biases.  Instead, to readily use the latching detector as a linear
preamplifier, we apply flux feedback in software to the QFP body loop, $\Phi^x_{\mathrm{qfp}}$ to keep the QFP at its balance
point where we have a 50\% probability of measuring it in either of its persistent current states after latching.  The required
$\Phi^x_{\mathrm{qfp}}$ is then a linear representation of the signal from the qubit.  At this balanced
population point, the QFP has minimal back-action on the qubit and can be used
to probe the magnitude of the circulating current of the qubit.  We use this mode of operation of the QFP in order to calibrate qubit parameters, as described in \cite{rhccjj}.
{\it Shift register readout:}
There is a well developed logic family based on QFPs \cite{gotoqfp}, and it is straightforward to design a QFP-based shift register that terminates in a single dc SQUID.  Such a readout architecture would require even fewer bias lines than our XY-addressable architecture.  We have successfully fabricated and tested an 8-QFP-long shift register driven with a three phase clock as a prototype of such a readout scheme.

\subsection{dc SQUID}\label{dcsquid}
The method for using a hysteretic dc SQUID for flux detection is discussed in \cite{harriscoupler} and more generally in
\cite{dcsquidreadout0,dcsquidreadout1}.  For the implementation discussed herein, current ramp times were typically $10-100 \mu s$ and
peak voltages from the dc SQUIDs after switching were typically kept to much less than the gap voltage of Nb to reduce heating on
chip.  The mutual inductance from the dc SQUID to the QFP was $M_{\mathrm{dc},\mathrm{qfp}}\sim 1.4$ pH.

\begin{figure}
\centering
\includegraphics[width=3.25in]{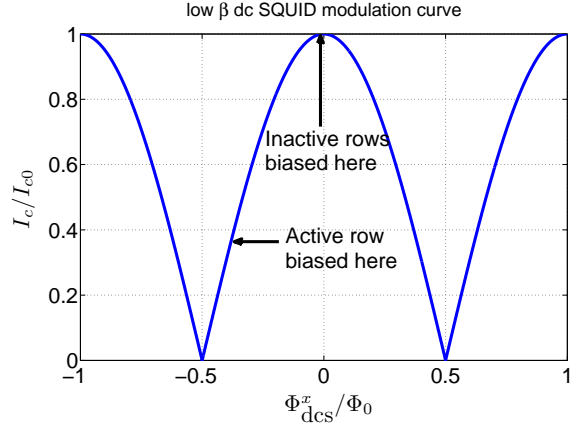}
\caption{Critical current $I_c$ vs. applied flux for a typical dc SQUID. Applying a signal via the $\mathrm{fb}_y$
  line allows a user to shift the operating point on the abscissa for a given row of dc SQUIDs.  Then
  when a current bias is applied to the $\mathrm{cb}_x$ line, only those SQUIDs whose critical current $I_c$
  has been suppressed will switch.}
\label{fig:dcsquidmodulation}
\end{figure}

Basic operation of the XY-addressable array of dc SQUIDs is presented
in figure~\ref{fig:waveforms}.  To read out the $i^{th}$ column and $j^{th}$ row QFP, we set the value of the current in the $\mathrm{fb}_j$ line in figure~\ref{fig:roschematic} to provide a flux bias of approximately 0.35 $\Phi_0$ into the dc SQUIDs on that row and $\mathrm{fb}_{k\ne j} = 0$ on all other rows.  This results in the selective suppression of the critical current for only the dc SQUIDs in row $j$ to $I_c = I_{c0} \cos(\pi 0.35) = 0.45 I_{c0}$.  Then, applying a current ramp on the column $\mathrm{cb}_i$ of maximum
amplitude $0.5 I_{c0}$ will cause the target dc SQUID to switch at a time that gives information about the flux coupled into dc SQUID $ij$ from its QFP. In
other words, the $\mathrm{fb}_y$ lines are used to row select by flux biasing all dc SQUIDs in the target row to a bias point where a
current ramp on the dc SQUID will trigger readout (see figure~\ref{fig:dcsquidmodulation}). The $\mathrm{cb}_x$ lines are used as column selects and all other current biases are kept at zero.  After we have registered an event, the current biases are reset to zero, and the process repeated for all combinations of $i$ and $j$.

For devices with less than a few thousand qubits it is feasible to provide enough analog lines to run an XY-addressable readout scheme as described.  Modifying this readout scheme by adding another control direction, for example an XYZ-addressable readout scheme where X and Z are summed together into the dc SQUID flux bias, changes the scaling of number of required wires to a cube root in the number of qubits, thereby allowing us practically scale up to reading tens of thousands of qubits.  If margins on the XYZ-addressable scheme are too low, then we can implement the QFP shift register method described earlier which uses a constant number of analog lines.  In that case we can use excess analog lines to decrease readout time or increase operating margins.

\subsection{Electronics}
The electronics to run the readout and qubit control were integrated into a compact rack system with
128 output channels, a fraction of which were devoted to readout.  The dedicated readout
channels had an amplifier, a tunable comparator chain and retriggerable timers in order to detect
multiple SQUID switching events per channel.  For the purposes of characterizing test circuits, we limited the readout sequence to a rate of approximately one dc SQUID switching event/100$\,\mu$s per channel.  Without modifying the line bandwidth (3 MHz), this could in principle be sped up to
roughly one event/$5\,\mu$s.  Further speed-up could be obtained by reading multiple columns simultaneously, but we have
some experimental evidence that rf isolation between columns needs to be improved as dc SQUIDs from different
columns can possibly interfere.  With further work, one can likely reach the bandwidth limitation of the wiring in the refrigerator.  For example, if a particular chip were to have 32 columns of dc SQUIDs and each line had a bandwidth of 3 MHz, one could obtain roughly a 50 Mbit/second readout rate.

\subsection{Power consumption}\label{designpower}
The readout process dissipates energy onto the chip whenever a dc SQUID is in the voltage state. An upper bound
on the total energy dissipated by the readout process can be estimated as follows: After the dc SQUID
switches to the voltage state, the large capacitance of the refrigerator wiring and filtering ($C = 2
nF$) starts to charge with a rate set by $\frac{dV}{dt} = I_{switch} / C$, typically $\sim 500 \mu V / \mu s$.
For simplicity, we ignore the first several nanoseconds after switching which are dominated by transients and ringing as the lines are not matched to the high impedance the dc SQUIDs present upon switching.  Until the dc SQUID reaches a significant fraction of the gap voltage there is little dissipation on chip.  For fast operation we typically restrict the charging time to less than two to three microseconds to keep the dc
SQUID from charging to the gap voltage and generating heat.  For the data presented in this paper, though, we
allowed many microseconds to pass (see figure~\ref{fig:dcsquidmultireads}).  The resulting dissipated energy is
$\Delta t \times 2.8 mV \times 1 \mu A \sim 28 \mathrm{fJ}$.  If we performed this same inefficient procedure
with a $100 \mu s$ readout cycle, we would obtain an average dissipated power during readout of $\sim 280
\mathrm{pW}$, which would be well within the capabilities of a standard dilution refrigerator to dissipate.

\section{Challenges}\label{challenges}

While we routinely fabricate and operate XY-addressable readout systems based upon the architecture described herein, there were some formidable challenges that had to be overcome in order to operate them to their full potential.  While the typical challenges associated with proper design of cryostat wiring, filtering, and shielding were encountered, we focus here on only those issues directly related to the readout scheme.

First, challenges related to fabrication yields are typically soft failures where one loses either a row or a column or even a single readout element.  Since the readout system discussed herein was typically a part of a large integrated circuit, it was possible to at least study large portions of a circuit even in the presence of a few fabrication errors.

Related to fabrication yield is the issue of critical current
variation across a chip.  For the dc SQUIDs, small variations of
critical current are not a significant issue as there is plenty of
range between OFF and ON switching currents in this XY-addressing
scheme.  The more frugal XYZ-addressable scheme described earlier,
though, would come at the expense of a reduced ON and OFF switching
current ratio in the presence of realistic critical current
variations.

Similar to critical current variation, flux offsets of each dc SQUID lead to reduced
ON/OFF margins as well.  In practice, this proved to be the key challenge to operating large scale XY-addressable
readout schemes in our laboratory.  We now use a combination of passive magnetic shielding and active magnetic field compensation to achieve fields below 1 nT perpendicular to the chip.  We
still typically end up with flux offsets of a few m$\Phi_0$ in QFPs and dc SQUIDs, which may be due to
local magnetic impurities on the wiring or in the magnetic shielding.  Cross-talk, both between off-chip and on-chip wiring, similarly reduces operating margins.

For the QFPs, asymmetry in the critical currents of their two Josephson junctions directly
create an apparent flux offset that competes with the qubit signal, as discussed previously.  Due to the threshold like latching, small
differences in critical current do very little, while large differences render the device unusable as the
qubit signal cannot overcome the asymmetry-induced flux bias acting on the QFP.  Typical requirements are less
than 2\% junction critical current asymmetry in the QFP in order for the device to be operable.

A further subtle effect, which is relevant to fast operation of the readout circuit, is thermally induced
hysteresis.  If one tries to perform experiments with a fast repetition rate (for tuned dc SQUID ramps, this
rate is typically 100 $\mu$s), then the chip temperature is affected by dc SQUID heating (see discussion in
\S\ref{readoutfidelity} below).  This dc SQUID heating depends quite strongly on when the dc SQUID escapes to
the voltage state, and therefore what signal is applied to the dc SQUID.  We have found that to measure any
temperature dependent phenomena, such as the transition width of a QFP or qubit, where one simultaneously
requires high speed and high accuracy measurements, we needed to temperature stabilize the mixing chamber of
the dilution refrigerator at 35-40 mK (much above the $<$ 10 mK base temperature of the refrigerators) and
take care to minimize the time the dc SQUID stays in the voltage state.

\section{Readout errors}\label{readoutfidelity}

Our research has shown that there  are two principle mechanisms response for imperfect readout in our XY-addressable readout.
The first mechanism is the thermal noise on the QFP, and the second mechanism is the
uncertainty in the switching time of the dc SQUID caused by the
stochastic nature of the quantum tunneling from the zero voltage state
to the voltage state\cite{mrt1}.  Each of these error mechanisms will be discussed below.

\begin{figure}
\centering
\includegraphics[width=3in]{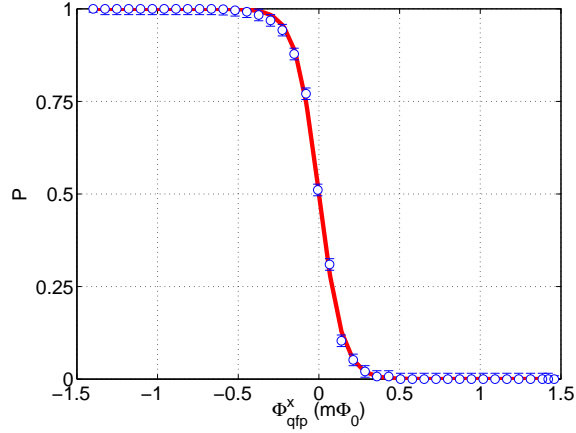}
\caption{Characterization of QFP flux sensitivity.  QFP state after latching as a function of flux added from
  the QFP flux bias line $\Phi^x_{\mathrm{qfp}}$.  The data (circles with error bars) fit to Eq.~\ref{eqn:tanh} (solid line), signifying that the detection is thermal noise dominated.  The transition width is 142$\,\mu\Phi_0$, which is to be compared to typical qubit signals of 10$\,$m$\Phi_0$.}
\label{fig:qfptransition}
\end{figure}

Examining the first error source, figure~\ref{fig:qfptransition} shows
the measured flux transfer curve of a typical QFP device, as obtained
by latching the QFP state in the presence of a constant flux through
the QFP flux bias line.  The data represent the probability of observing one of the QFP persistent current states $P$ versus $\Phi^x_{\mathrm{qfp}}$.  Note that the data yield a smoothed step function with a width $\sim 142 \mu\Phi_0$.  This finite width is due to imperfect initialization of the QFP state.

To understand the source of this QFP transition width, it suffices to
analyze the dynamics of the two lowest QFP energy levels as the
barrier $\Delta U_{\mathrm{qfp}}$ between flux states is raised (as in
figure~\ref{fig:qfppotentialenergy}) in the presence of a thermal
bath.  As the barrier is raised, the tunneling energy
$\Delta_{\mathrm{qfp}}$ between localized flux states is reduced to
zero. Once $\Delta U_{\mathrm{qfp}}$ is high enough and
$\Delta_{\mathrm{qfp}}$ low enough the system is no longer able to
maintain a state that is in thermal equilibrium.  We estimate this dynamical
freeze-out to occur when $\Delta_{\mathrm{qfp}}/h \sim 1$ MHz, given
our QFP annealing times (time taken to raise the QFP
barrier) of roughly 1 $\mu$s.  The dynamics of the QFP as $\Delta_{\mathrm{qfp}}$ approaches this small value are well described by the incoherent tunneling
picture studied in Refs. \cite{aminaverin,harrismrt}.  The value of the QFP persistent
current, calculated from extracted device parameters at that point, is
$I^p_{\mathrm{qfp}} \sim 4.5\pm 0.5 \mu$A.  We can then 
convert the abscissa in figure~\ref{fig:qfptransition} from flux to
energy (2 $\times\Phi^x_{\mathrm{qfp}}\times$ $I^p_{\mathrm{qfp}}$).
By fitting the QFP population as a function of $\Phi^x_{\mathrm{qfp}}$
to the expected functional form in the limit
$\Delta_{\mathrm{qfp}}\rightarrow 0$,
\begin{equation}
\label{eqn:tanh}
P=\frac{1}{2}\left[1-\tanh\left(\frac{2\Phi^x_{\mathrm{qfp}}I^p_{\mathrm{qfp}}}{2k_BT}\right) \, \right],
\end{equation}

\noindent we extract an effective $T=47\,$mK for the QFP and its local bath.  For comparison, the refrigerator was stabilized at
approximately 40 mK.  If we decrease the cooling time after the dc
SQUID switching event before the QFP latch in the next measurement
frame we can characterize the cooling of the QFP after the dc SQUID
heating event:
\begin{center}\begin{tabular}{|c|c|}
\hline
Cooling time ($\mu$s) & Temperature (mK)\\
\hline
300 & $75\pm7$ \\
1000 & $62\pm6$ \\
2000 & $55\pm5$ \\
4000 & $47\pm5$ \\
\hline
\end{tabular}\end{center}
When operated with dc SQUID current biases tuned to minimize heating,
or with sufficient cooling time such as the effective 47 mK data shown
in figure~\ref{fig:qfptransition}, we obtain a corresponding flux
sensitivity of the QFP of $k_BT/I^p_{\mathrm{qfp}}=142\,\mu\Phi_0$.  For comparison, the qubit
signal coupled into the QFP is of order 10 m$\Phi_0$.  This results in
practically zero error probability for the latch.

The QFP flux uncertainty (142 $\mu \Phi_0$) is significantly better than that of its accompanying dc SQUID, which was estimated to be 1.8$\,$m$\Phi_0$ from the width of its switching distribution.  This dramatic difference in sensitivity arises from the fact that dc SQUID noise results from the stochastic nature of quantum tunneling from the zero voltage to finite voltage state with an equivalent temperature of several hundred mK, while the QFP annealing step leads to a thermal uncertainty near the refrigerator temperature of 40 mK.

\begin{figure}
\includegraphics[width=3in]{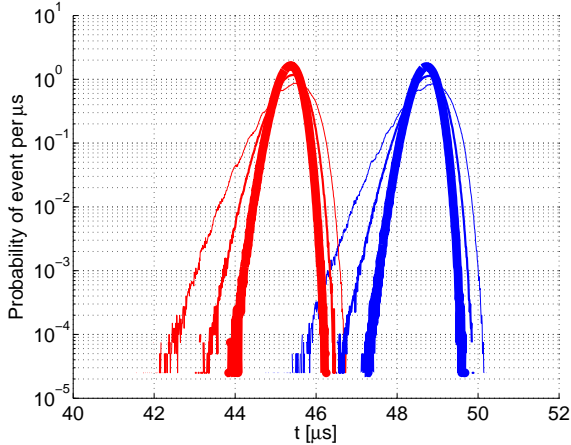}
\caption{Reducing readout errors with repeated dc SQUID sampling of
  the QFP state.  The dc SQUID is interrogated with a 50 $\mu$s long
  linear current bias ramp. The red and blue curves correspond to
  different initialized flux states of the qubit (which is then
  adiabatically transferred to the QFP).  The three lines (from thin
  to thick) correspond to 1, 2, and 4 averaged reads of the dc SQUID.
  Once 4 reads are performed we see no overlap in the data set, which
  was 4 million points.  From the thick lines we extract an error
  probability of $10^{-6}$ that is quoted in the text.  The small
  steps in the switching histograms are due to the resolution of the
  room temperature electronics.  The measurements shown in this plot
  were obtained from one readout on an 8-qubit
  chip.}\label{fig:dcsquidmultireads}
\end{figure}

Focusing now on the second error mechanism, figure
\ref{fig:dcsquidmultireads} shows measured switching histograms of a
single dc SQUID for the case when its respective qubit is first
initialized in one state (red) and then in the other (blue).  While
the dc SQUID switching distributions are characteristic of quantum
tunneling and thus asymmetric, we still can characterize the dc SQUID
sensitivity by the standard deviation of the switching histogram.  The
abscissa can be converted from time to flux by direct measurement of
the shift in switching time as a known amount of flux is applied to
the dc SQUID.  The resulting dc SQUID sensitivity is 1.8 m$\Phi_0$ per
read, and the signal from the QFP latch is 11.9 m$\Phi_0$, resulting
in a measured error probability of the dc SQUID in reading the QFP as
small as 0.01.  This is a reasonably low error probability, but we
require even lower error rates in order to implement a large scale AQO
processor.  It is trivial to decrease this dc SQUID limited error by
taking more samples of the dc SQUID switching time since the QFP
maintains its state between these samples.  The results of this
repeated reading of the dc SQUID on its resolving power are shown in
figure ~\ref{fig:dcsquidmultireads}.  After 2 reads, the dc SQUID
sensitivity increases to 1.3 m$\Phi_0$ and after 4 reads to 0.9
m$\Phi_0$.  At 4 reads we obtain an error probability of $10^{-6}$.
This error probability includes qubit initialization errors, QFP
errors, and dc SQUID errors.  This behaviour, while varying slightly
quantitatively, was similar for all 8 readouts on the particular test
chip used to generate these data.

\begin{figure}
\centering
\includegraphics[width=3.25in]{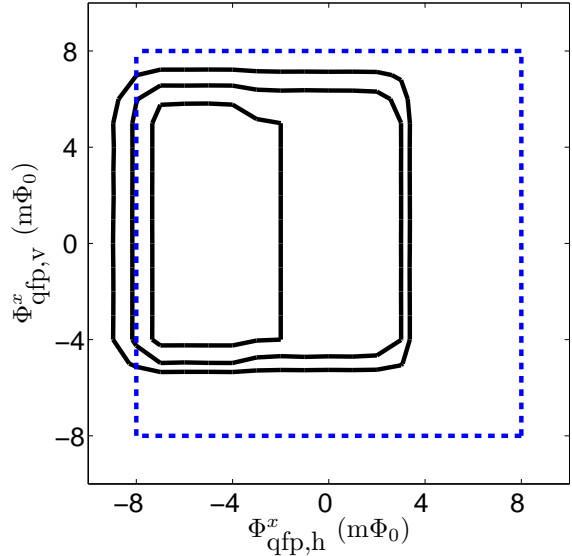}
\caption{Measured QFP margins on a 128-qubit circuit.  Three concentric
  black rings represent contours where the system returned 1/128,
  4/128 and 9/128 failures, respectively, from inside to outside.  The dashed blue box indicates the designed operating margins.}
\label{fig:readoutqfpmargins}
\end{figure}

To demonstrate the performance of a large-scale XY-addressable readout system, we
present the results of a system level margin measurement from a
128-qubit chip in figure ~\ref{fig:readoutqfpmargins}.  This
particular chip had two flux biases to provide $\Phi^x_{\mathrm{qfp}}$
to the 128 QFPs: The two lines each biased 64 QFPs, with the two
groups partitioned depending upon the physical orientation of the QFP
body and denoted as ``horizontal'' or ``vertical''.  The reader is referred to \cite{PhysRevB.82.024511} for a description of the 128-qubit chip physical layout.  We then studied
the error probability of the readout system as a function of
horizontal $\Phi^x_{\mathrm{qfp,h}}$ and vertical
$\Phi^x_{\mathrm{qfp,v}}$ QFP biases.  In order to render the readout
test independent of the details of quantum annealing, we prepared the
128 qubits into a known state by setting all of the inter-qubit
couplings to zero \cite{harriscoupler}, biasing all qubits using their
local flux sources \cite{mwjpmm} hard to one side, and then raising
their tunnel barriers.  This placed each flux qubit into a known state
with certainty.  We then read the state of each qubit, as described
above using a single dc SQUID switching measurement.  This sequence
was repeated 128 times for every point on a 2-dimensional grid in
$(\Phi^x_{\mathrm{qfp,h}},\Phi^x_{\mathrm{qfp,v}})$-space.  The
resultant contours of constant error are shown in figure
\ref{fig:readoutqfpmargins} as black lines, as well as the designed
margins as a blue dashed line.  Inside the 1 error contour we observed
no errors out of 128 reads of the entire 128 qubit system.  These
measurements alone confirm that the system level error rate can be
less than $1/128\approx 0.008$ when appropriately biased.  The offset
of the measured margins in the horizontal direction was due to a
single QFP with a flux offset, either from Josephson junction
asymmetry or flux pinning, of $-10\,$m$\Phi_0$.  When this particular
readout failed due to the inability of the qubit signal to straddle
that QFP's degeneracy point, it contributed 1/128 to the total failure
rate, thus dominating the shape of the inner most contour.  The next
margin contour then required a much larger horizontal flux bias shift
to cause a second QFP to fail repeatedly.  The slow increase in
failure rates is due to a several m$\Phi_0$ random distribution of
flux offsets in the QFPs.

In order to better quantify the performance of the 128 qubit readout
system, we biased the QFPs to an operating point near the center of
the margin plot,
$(\Phi^x_{\mathrm{qfp,h}},\Phi^x_{\mathrm{qfp,v}})=(-5,0)\,$m$\Phi_0$,
and performed 65536 measurements in the manner cited above.  We
observed 5 errors out of 65536 measurements, which then provided an
estimated system level error probability of $5/65536\approx 8\times10^{-5}$.  The
five errors were attributed to different readouts and were probably
due to the stochastic nature of the dc SQUID switching.  From these
results, we can crudely estimate the per qubit error to be on the
order of $5/(65536\times 128)\approx 6\times 10^{-7}$ for the 128
qubit chip.  This could be improved upon by simply using repeated dc
SQUID reads, as was done with the 8 qubit chip, and the final limit
upon the readout error probability will likely be determined by outlier events not captured in the preceding discussions.

\section{Conclusions}
We have described and implemented a scalable XY-addressable readout scheme that is
suitable for a superconducting adiabatic quantum optimization system
and measured the readout error rate and margins of an 8- and 128-qubit
system, respectively.  We conclude that the single qubit readout error rate can be
made less than $10^{-6}$ using the QFP enabled architecture, which
makes readout errors a non-issue for large scale integrated circuits
of the types considered herein.  We have experimentally demonstrated a
system level error rate of $8\times10^{-5}$ using a 128 qubit circuit.
We do not expect any significant challenges in further scaling this
architecture to about one thousand qubits.  Finally, we have briefly
discussed both an XYZ-addressable and a QFP shift register based solution that
would facilitate scaling up to read even larger numbers of qubits on a
single chip.

\section*{Acknowledgments}

The authors would like to thank F. Brito, B. Bumble, D. Bruce, E. Chapple, P. Chavez, V. Choi,
F. Ciota, C. Enderud, A.~K.~Fung, S. Govorkov, S.~Han, J. P. Hilton, A. Kaul, A. Kleinsasser,
D. Klitz, G. Lamont, P. Lumsden, F. Maibaum, R. Neufeld, T. Oh. L. Paulson, I. Perminov, C. Petroff,
D. Pires, C. Rich, P. Spear, A. Tcacuic, M. Thom, S. Uchaikin, M. Wang, and A. B. Wilson for work on
cryogenics, electronics, fabrication, filtering, and magnetic shielding.

\section*{References}
\bibliographystyle{unsrt}
\bibliography{readoutsystem}

\end{document}